\documentclass[12pt,preprint]{aastex}
\usepackage[loose,ugly]{units}
\usepackage[onecolumn]{emulateapj5}
\RequirePackage{amssymb}
\RequirePackage{amsbsy}
\newcommand{\bvec}[1]{\ensuremath{\boldsymbol{#1}}} 
\newcommand{\divr}[1]{\bvec{\nabla \cdot {#1}}} 

\newcommand{\nuclei}[2]{\ensuremath{\mathrm{^{#1}#2}}}
\newcommand{\Hyd}{\ensuremath{\mathrm{H}}}

\newcommand{\Hethree}{\nuclei{3}{He}}
\newcommand{\C}{\nuclei{12}{C}}
\newcommand{\Cth}{\nuclei{13}{C}}
\newcommand{\N}{\nuclei{14}{N}}
\newcommand{\Nth}{\nuclei{13}{N}}
\newcommand{\Oxygen}{\nuclei{16}{O}}

\newcommand{\D}{\ensuremath{\mathcal{D}}}

\newcommand{\J}{\ensuremath{\bvec{J}}}

\newcommand{\JDNB}{\ensuremath{\bvec{J}_\mathrm{DNB}}}

\newcommand{\Msun}{\ensuremath{M_\odot}}

\newcommand{\cen}{\mbox{Cen~X-4}}

\newcommand{\CM}{\ensuremath{\,\unit{cm}}}

\newcommand{\tgccm}{\ensuremath{\,\mathrm{\unit{g}\,\unit{cm^{-3}}}}}
\newcommand{\gsqcm}{\ensuremath{\,\unitfrac{g}{cm^2}}}
\newcommand{\tgsqcm}{\ensuremath{\,\mathrm{\unit{g}\,\unit{cm^{-2}}}}}
\newcommand{\yrs}{\ensuremath{\,\unit{yrs}}}
\newcommand{\Kelvin}{\ensuremath{\,\unit{K}}}
\newcommand{\Gauss}{\ensuremath{\,\unit{G}}}
\newcommand{\sqcmg}{\ensuremath{\,\unitfrac{cm^2}{g}}}

\newcommand{\secs}{\ensuremath{\,\unit{s}}}

\shorttitle{Diffusive Nuclear Burning in NS Envelopes}
\shortauthors{Chang, P. \& Bildsten, L.}

\begin{document}

\title{Diffusive Nuclear Burning in Neutron Star Envelopes}


\author{Philip Chang} 
\affil{Department of Physics, Broida Hall, University of California,
Santa Barbara, CA 93106; pchang@physics.ucsb.edu}
\and
\author{Lars Bildsten}
\affil{Kavli Institute for Theoretical Physics and Department of
Physics, Kohn Hall, University of California, Santa Barbara, CA 93106;
bildsten@kitp.ucsb.edu}


\begin{abstract}
  
  We calculate the rate of hydrogen burning for neutron stars (NSs)
  with hydrogen atmospheres and an underlying reservoir of nuclei
  capable of proton capture. This burning occurs in the exponentially
  suppressed diffusive tail of H that extends to the hotter depths of
  the envelope where protons are rapidly captured. This process, which
  we call diffusive nuclear burning (DNB), can change the H abundance
  at the NS photosphere on timescales as short as $10^{2-4}$ years. In
  the absence of diffusion, the hydrogen at the photosphere (where
  $T\approx 10^6 \Kelvin$ and $\rho\sim 0.1 \tgccm$) would last for
  far longer than a Hubble time. Our work impacts the understanding of
  the evolution of surface abundances of isolated NSs, which is
  important to their thermal spectrum and their effective
  temperature-core temperature relation. In this paper, we calculate
  the rate of H burning when the overall consumption rate is
  controlled by the nuclear timescales, rather than diffusion
  timescales. The immediate application is for H burning on
  millisecond radio pulsars and in quiescence for the accreting NS Cen
  X-4. We will apply this work to young radio pulsars and magnetars
  once we have incorporated the effects of strong $B>10^{12} \Gauss$
  magnetic fields.

\end{abstract}


\section{Introduction}

The increasing number of observations of isolated neutron star (NS)
atmospheres (see Pavlov, Zavlin \& Sanwal 2002 for a recent review)
and resulting constraints on their surface composition has highlighted
the need to consider the role of nuclear physics during their cooling
phase. The initial composition of the outer layer is neither known nor
constrained by the theory of supernova explosions since the amount of
matter needed to affect the outer envelope is small ($\sim 10^{-20}
\Msun$ for the photosphere). This much material can easily fallback
and contaminate the outer atmosphere (\citealt{woosley.95}). Even if
the fallback material consists of heavy elements, spallation (see
Bildsten, Salpeter \& Wasserman 1992) will occur at some stage in the
fallback, creating a plethora of lighter elements that rapidly
gravitationally separate.  Current evidence suggests that young
pulsars ($< 10^{4-5}\yrs$) possess photospheres of either hydrogen or
helium, whereas older pulsars ($> 10^5 \yrs$) appear to have envelopes
of heavier, more uniformly opaque elements (Yakovlev, Kaminker \&
Gnedin 2001).  This may indicate an evolution of the outer envelope on
the timescale of $10^{4-5} \yrs$, something we would like to
understand better.

As part of our ongoing work on this problem, we present here a
mechanism of nuclear processing of the outer envelope which we call
diffusive nuclear burning (DNB). The basic picture is simple and was
alluded to by \citet{chiu.64} as a mode of nuclear burning.
\citet{rosen.68} carried out the initial calculations of the effect
for surface temperatures in the range of $(5-10)\times 10^6
\Kelvin$. Their work was in the limit where the diffusion time to
larger depths was the rate-limiting step.

\begin{figure}[tbp]
  \epsscale{0.75}
  \plotone{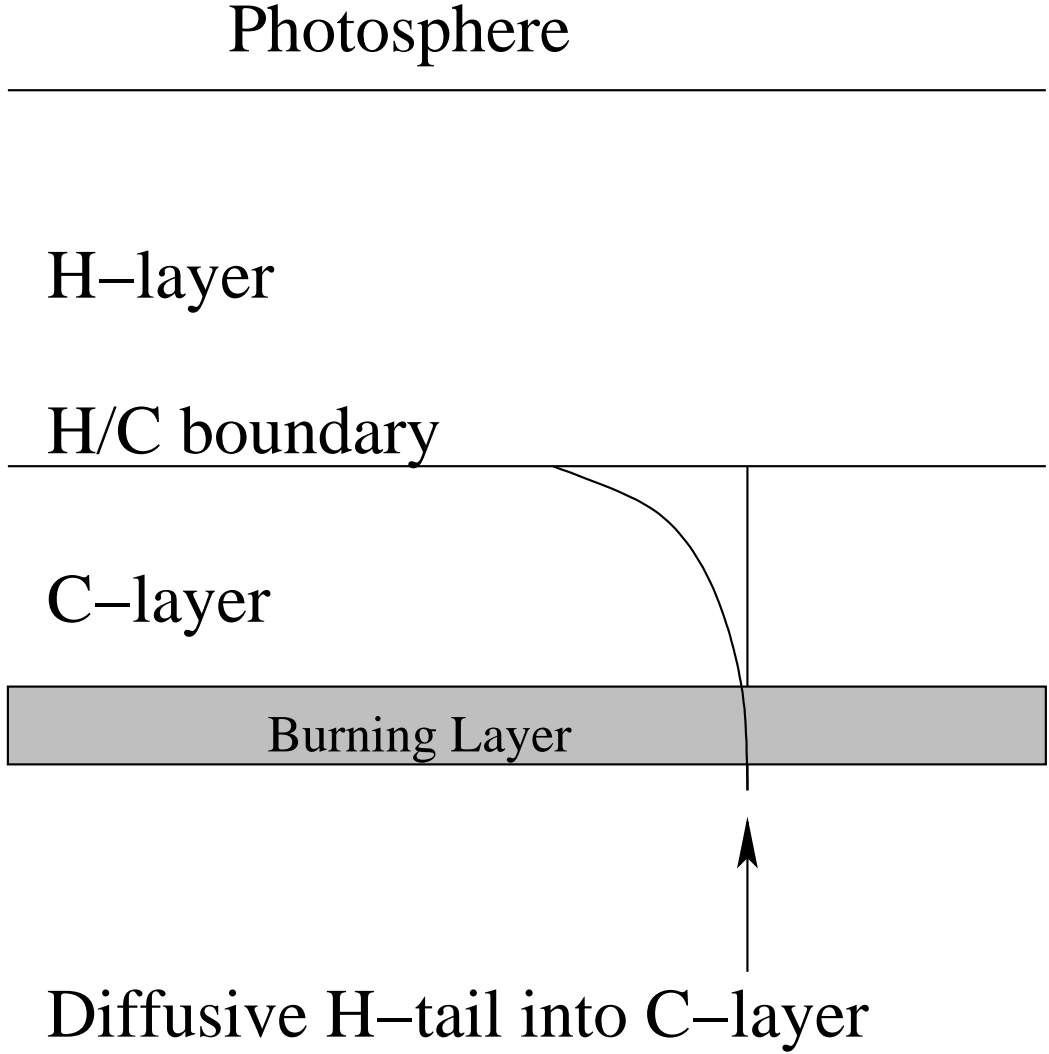}
  \caption{Diagram of a H/C envelope in diffusive equilibrium.
    The diffusive tail of hydrogen extends deep into the
    carbon, reaching temperatures where the hydrogen 
rapidly captures onto carbon.}
  \label{fig:envelope_diagram}
\end{figure}

In this paper, we will consider much lower values for the surface
temperature, where diffusion is not the limiting step. Consider a NS
envelope that consists of hydrogen above a layer of carbon (or any
other proton capturing material) in diffusive equilibrium (see
illustrative Figure \ref{fig:envelope_diagram}). For simplicity, we do
not calculate the effect of an intervening helium layer in this paper,
though we estimate its possible impact in the conclusions. As long as
there has been adequate time to reach a diffusive equilibrium, the
separation between the hydrogen and carbon is not strict and a
diffusive tail of hydrogen penetrates deep into the carbon layer. The
temperature rises with depth and can increase by 1-2 orders of
magnitude 10 meters beneath the photosphere.  At this location, the
lifetime of a proton is very short. The convolution of the increasing
nuclear burning rate and the decreasing hydrogen abundance creates a
burning layer at this depth, where hydrogen burning peaks. As hydrogen
burns, the diffusive tail is driven slightly away from diffusive
equilibrium and a diffusive current is set up, \JDNB.  We will always
work in the limit where \JDNB\ is far less than the maximum current
that can be supported due to diffusive processes so that diffusive
equilibrium remains an excellent approximation.

We find that, even when the photospheric temperatures and densities
are so low that the local nuclear lifetime of a proton is in excess of
a Hubble time, DNB can proceed on a much faster timescale.  In this
paper we make some simplifying assumptions that prohibits a direct
application to radio pulsars, and thus the extension of the physics to
radio pulsars and magnetars will be left to future work. For example,
we require a magnetic field low enough so that there is a negligible
effect on the temperature profile; $B < 10^9 {\rm G}$
(\citealt{ventura.potekhin.01,potekhin.yakovlev:magnetized.envelopes}).
However, our work is applicable to some X-ray transients in
quiescence, where the outer envelope consists of pure hydrogen, even
though the depth of the hydrogen layer is unknown.

This paper is structured as follows.  In \S~\ref{sec:microphysics} we
present the basic equations and microphysics governing the thermal and
compositional structure of a NS envelope in diffusive equilibrium.
The derivation of the electric field strength from the envelope
structure equations is in \S~\ref{sec:electric_field}. Using the
derived electric field strength, we numerically solve the thermal and
compositional structure of the NS envelope. The analytic solution to
the NS envelope structure is given in
\S~\ref{sec:approx_temp_comp}. In \S~\ref{sec:hydrogen_burning} we
discuss the relevant hydrogen burning reactions that drive the NS
envelope evolution, present numerical and analytic solutions for
\JDNB, and apply our theory to the X-ray transients, specifically
\cen. We summarize our findings in \S~\ref{sec:discussion}.

\section{Atmospheric Structure and Microphysics}\label{sec:microphysics}

We only consider the top $10^4 \ {\rm cm}$ of the envelope, where the
density is $<10^{10} \tgccm$.  Since the thickness of the envelope is
much smaller than the NS radius ($R \approx 10$ km), we presume a
plane parallel atmosphere with a constant downward gravitational
acceleration, $g = GM/R^2$. We neglect all relativistic corrections.
The structure of the NS envelope is determined by the equations of
hydrostatic balance,
\begin{eqnarray}\label{eq:hb}
\frac {dP_i} {dr} &=& -n_i \left( A_i m_p g - Z_i e E\right), \\
\frac {dP_e} {dr} &=& -n_e\left(m_e g + e E\right),
\end{eqnarray}
where $P_i$, $n_i$, $A_i$, $Z_i$ are the pressure, number density,
atomic number and charge of the $i$'th ion species and $E$ is the
upward pointing electric field.  The thermal structure is determined
by the heat diffusion equation for a constant flux, $acT_e^4/4$, where
$T_e$ is the effective temperature,
\begin{equation}
\frac {dT} {dr} = -\frac {3 \kappa \rho} {16 T^3}T_e^4, 
\label{eq:flux}
\end{equation}
where $\kappa$ is the opacity. Finally, we demand charge neutrality,
\begin{equation}\label{eq:charge_neutrality}
n_e = \sum_i  Z_i n_i. 
\end{equation}
Summing the pressure equations with the charge neutrality constraint
recovers the familiar form of hydrostatic balance, $dP/dr = -\rho g$.
 
In the outermost layers of a NS envelope, the nuclei are fully ionized
and the radiative opacities are set by free-free absorption and
Thomson scattering,
\begin{equation}
\kappa_{\rm Th} = \frac {n_e \sigma_{Th}} {\rho} \approx \frac {0.4}
{\mu_e} \sqcmg,
\end{equation}
where $\mu_e = A/Z$ is the electron mean molecular weight. 
We calculate the free-free opacity by adopting the formalism used in
Schatz et al. (1999),
\begin{equation}
\kappa_{\rm ff} = 7.53 \times 10^6 \frac {\rho_5} {\mu_e T_6^{7/2}}
\sum \frac {Z_i^2 X_i} {A_i} g_{\rm ff}(Z_i, T, n_e) \sqcmg,
\label{eq:kramers_opacity}
\end{equation}
where $T_6=T/10^6\Kelvin$, $\rho_5=\rho/10^5\tgccm$, $X_i$ is the mass
fraction, and $g_{\rm ff}$ is the Gaunt factor.  For our numerical
calculations, we calculate the Gaunt factor using the \cite{schatz.99}
fitting formula [given by their equation (A2)].  For our analytic
calculations we set $g_{\rm ff}=1$.

At very high densities, the dominant mode of heat transport is
conduction by degenerate electrons.  For strongly coupled Coulomb
plasmas (SCCP), the electron thermal conductivity is determined by
electron-ion scattering as given by the Wiedemann-Franz law
(\citealt{ventura.potekhin.01}; Brown, Bildsten \& Chang 2002).
\begin{equation}
K_c = \frac {\pi^2 k_B^2 T n_e \tau_{K}} {3 m_e^*},
\end{equation}
where $m_e^* = \epsilon_{\rm F}/c^2$ is the effective electron mass,
$\epsilon_{\rm F}$ is the electron Fermi energy, and $\tau_{K}$ is the
effective relaxation time. For weak magnetic fields ($B < 10^9$G), we
use the thermal conductivity of electrons in SCCP given by Baiko et
al. (1998) with analytic formulae given by Potekhin et al. (1999).  In
the liquid metal approximation, \citet{brown.02} showed that this
formalism agrees with the calculations of \citet{urpin80} and Itoh et
al. (1983).

The transition of the envelope from primarily radiative to primarily
conductive takes place over a narrow region called the sensitivity
strip (\citealt{ventura.potekhin.01}).  Heat transport by radiation
and conduction is of the same order in this region.  The sensitivity
strip also plays the dominant role in determining the temperature
profile of the NS envelope (as discussed in
\S~\ref{sec:approx_temp_comp}, also see
\citealt{ventura.potekhin.01}).  In the sensitivity strip, the SCCP is
always in the liquid metal regime, whose conductivity is fairly well
modeled (\citealt{brown.02, ventura.potekhin.01}).  Thus, we
are confident in our results for the thermal profile.

The electron equation of state transitions from an ideal gas EOS to a
degenerate EOS.  To model this transition in our numerical solutions,
we have used the \citet{paczynski.83} equation of state.  In our
analytic calculations, we solve our envelope for the two different
limits and introduce a free parameter $\Psi$ to join them at a point
defined by
\begin{equation}\label{eq:degenerate_cond}
  \frac {P_{\rm e, deg}} {P_{\rm e, nondeg}} \equiv \Psi,   
\end{equation}
where $P_{\rm e, deg}$ and $P_{\rm e, nondeg}$ are the nonrelativistic
degenerate and nondegenerate electron equations of state respectively.
For this calculation we take $\Psi = 1.5$, which gives a reasonable
approximation to the full numerical solution in the region of the envelope
where DNB occurs.

\section{Elemental Distribution and Electric Field Strength}\label{sec:electric_field}

The composition of the envelope in diffusive equilibrium is found by
simultaneously solving the equations of hydrostatic balance for each
species and the flux equation. This approximation is excellent once
enough time has passed for the diffusive equilibrium to be
established (\citealt{brown.02}).

We begin by calculating the local electric field.  In early studies of
white dwarfs and NSs (\citealt{rosen.68, fontaine.79}), the
electric field was presumed to be that for an ideal gas of ions and
electrons,
\begin{equation}
  eE = \frac A {Z + 1} m_p g. 
  \label{eq:rosseland_E}
\end{equation}  
However, in the degenerate case where $P_e\approx P$, the field is
\begin{equation}
  eE = \frac A Z m_p g.
  \label{eq:degenerate_E}
\end{equation}
These results are only valid in certain limits, where the plasma
consists of one dominant species and where the gas is either very
degenerate or very non-degenerate.  For multi-ion plasmas and for
partial degeneracy, these simple approximations break down. For a more
general solution, we calculate the electric field from the ion and
electron equations of hydrostatic balance (eq. [\ref{eq:hb}]), and the
flux equation (eq. [\ref{eq:flux}]). We presume an ideal gas equation
of state for the ions and an arbitary electron equation of
state. Finally we drop the $m_eg$ term.

Let us first consider the isothermal case. We first expand $\partial
P_e/\partial r$,
\begin{equation}
  \frac {\partial n_e} {\partial r} = -\left(\frac {\partial P_e}
  {\partial n_e}\right)^{-1} n_eeE ,  \label{eq:n_e_HB}
\end{equation}
and then use the ideal gas equation of state for the ions:
\begin{equation}
  \frac {\partial n_i} {\partial r} = -\frac {n_i} {kT} ( A_im_pg -
  eZ_iE).  \label{eq:n_ion_HB}
\end{equation}
Multiplying each ion equation (eq. [\ref{eq:n_ion_HB}]) by $Z_i$ and
subtracting their sum from the electron equation (eq. [\ref{eq:n_e_HB}]) and
presuming charge neutrality (eq. [\ref{eq:charge_neutrality}]) gives
\begin{equation}
  eE = \frac {m_pg\sum n_i A_i Z_i} { \sum n_iZ_i^2 + n_e k_B T 
    \left({\partial P_e}/{\partial n_e}\right)^{-1}}.
\end{equation}
For the nonisothermal case, we get an additional term in the expansion
of $\partial P_e/\partial r$ and $\partial P_i/\partial r$, e.g.
\begin{equation}
  \frac {\partial n_e} {\partial r} = -\left(\frac {\partial P_e}
{\partial n_e}\right)^{-1} \left(n_eeE - \frac {\partial P_e}
{\partial T} \frac {\partial T} {\partial r}\right),
\label{eq:ne_HB}
\end{equation}
and the electric field becomes
\begin{equation}
  eE = \frac {\sum n_i Z_i \left(A_im_p g + k_B \left({dT}/{dr}\right)\right) -
\left( {\partial P_e}/{\partial T}\right)\left({\partial P_e}/{\partial
n_e}\right)^{-1} \left({dT}/{dr}\right)} { \sum n_iZ_i^2 + n_ek_B T \left(
{\partial P_e}/{\partial n_e}\right)^{-1}}.
\end{equation}
This result agrees with the results of Macdonald, Hernanz \& Jose
(1998) and \citet{althaus.2000}, but we have decoupled our solution of
the electric field from the drift velocity and diffusion coefficient.
Hence, the electric field is simply a function of local parameters.
For the purposes of numerical calculation we use the
\citet{paczynski.83} electron equation of state.  For our analytic
calculations, we use either the ideal gas or the nonrelativistic
degenerate electron equation of state depending on the local
conditions of the plasma.

\begin{figure}[tbp]
  \epsscale{0.80}
  \plotone{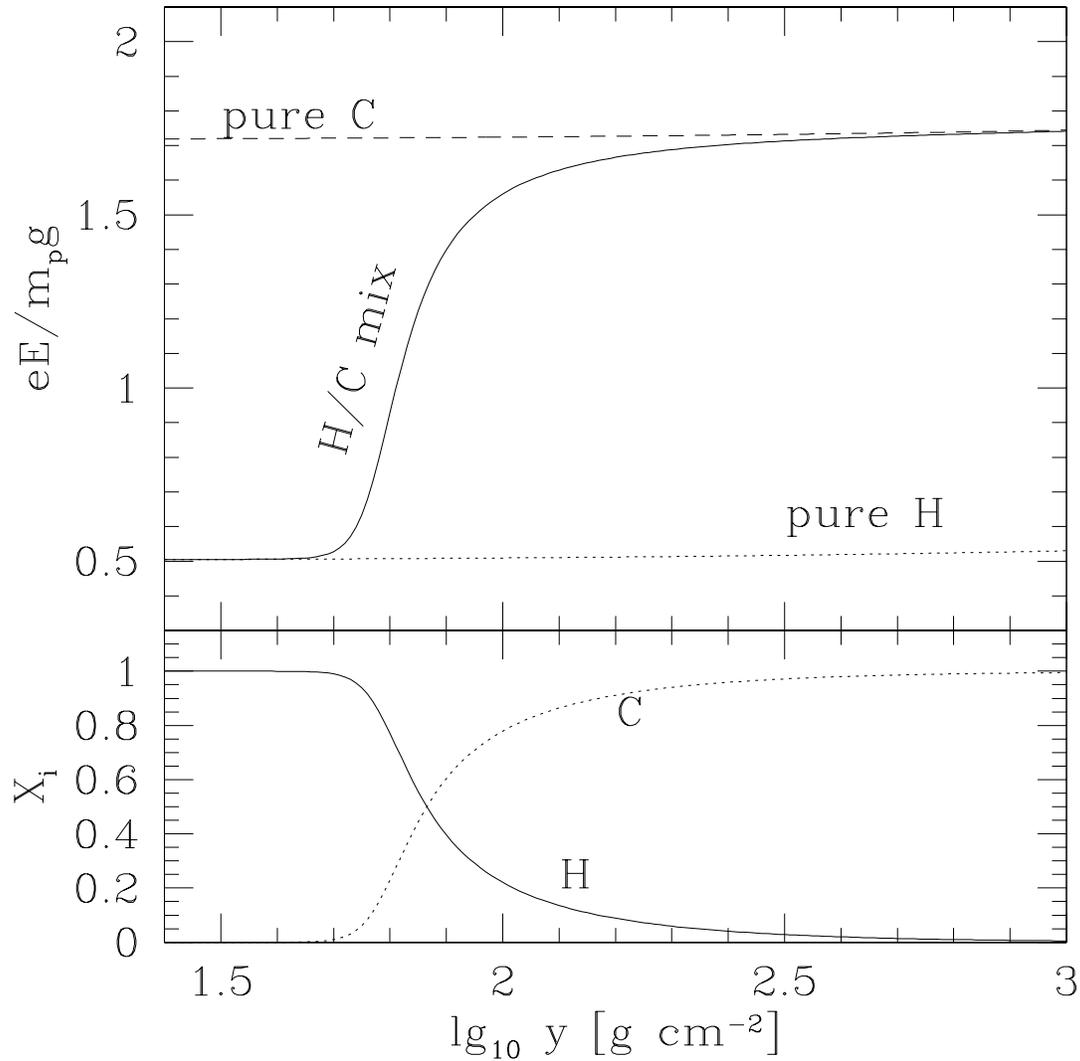}
    \caption{Electric field strength in an atmosphere in diffusive
      equilibrium.  The electric field is shown for a total column of
      hydrogen of $y_H = 100 \tgsqcm$ and $T_e = 8 \times 10^5
      \Kelvin$.  The electric field changes from the value for a pure
      hydrogen atmosphere to that for a pure carbon atmosphere.  The
      bottom graph shows the mass fraction of the two components,
      $X_i$, as a function of column.  The variation in $X_i$ traces
      out the variation in the electric field.  The electric field
      varies from one limit to another over a zone of order of a scale
      height, which agrees with the results of \citet{alcock.80a}.}
    \label{fig:electric_field}
\end{figure}

\begin{figure}[tbp] 
  \epsscale{0.80}
  \plotone{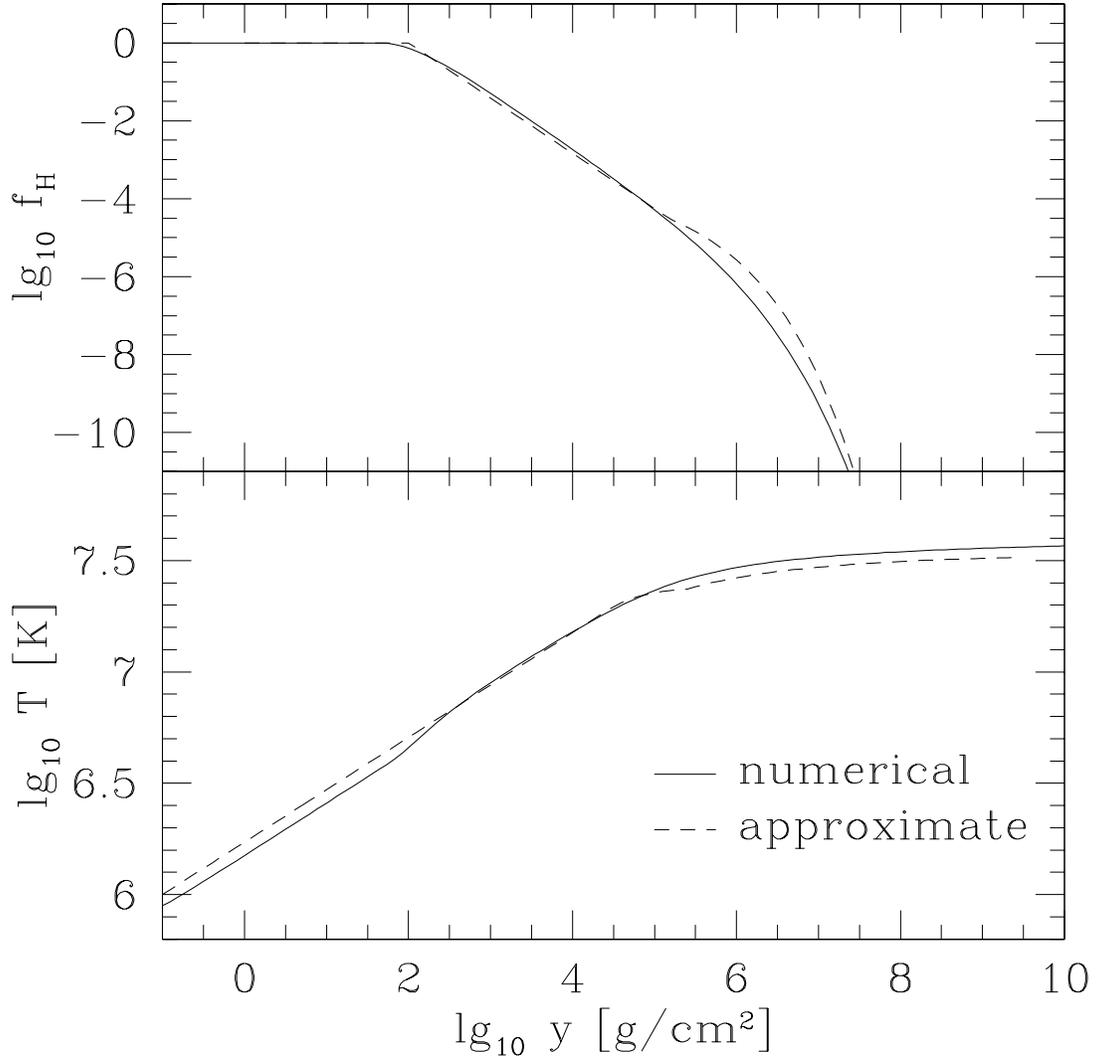}
  \caption{Thermal structure and composition for a NS in
    diffusive equilibrium with $T_e = 8 \times 10^5 \Kelvin$, $g_{14}
    = 2$, and $y_H = 100 \tgsqcm$.  The solid line is the numerical
    solution for the temperature and hydrogen number fraction, $f_H =
    n_H/n_{\rm tot}$, as a function of column depth.  The dashed line
    is the approximate solution for the temperature and hydrogen
    number fraction, $f_H$. The approximate analytic solution for
    temperature is given by equation (\ref{eq:thermal_nd_sol}) and
    (\ref{eq:temp_degenerate_sol}).  The approximate solution for
    $f_H$ is given by equations (\ref{eq:sol_conc_nondeg}) and
    (\ref{eq:comp_deg_sol}) with $\Psi = 1.5$.}
    \label{fig:col_vs_conc_T}
\end{figure}

At this point, it is helpful to illustrate a solution for the
particular case of a \Hyd/\C\ envelope.  In Figure
\ref{fig:electric_field}, we plot the electric field of the layer as a
function of column depth, $y = P/g$.  We integrate from the
photosphere with a background of \Hyd.  At a certain depth we
introduce a low carbon abundance and continue integrating inward.  As
expected, the electric field approximates that of the value given by
equation (\ref{eq:rosseland_E}) at the extreme limits where one ion
dominates over the other.  The slight discrepancy between the value
give by equation (\ref{eq:rosseland_E}) at the extreme limits is due
to degeneracy effects from the \citet{paczynski.83} fit of the
electron equation of state.  The flux and structure equations can now
be solved self consistently to model any arbitrary envelope in
diffusive equilibrium. In Figure \ref{fig:col_vs_conc_T}, we plot the
solution for a \Hyd/\C\ envelope.  We plot the composition
profile in terms of the concentration of hydrogen, $f_H = n_H/n_t$,
where $n_t$ is the total number of ions.

\section{Approximate Solutions for Temperature and Composition}\label{sec:approx_temp_comp}

The temperature and composition profiles can be represented well with
approximate analytic solutions that illuminate the physics.
\citet{hernquist.applegate:NS_analytic_calculation} and
\citet{ventura.potekhin.01} have presented analytic thermal
calculations of the NS envelope, but their solutions are too complex
for our analytic estimates.  Hence, we present simplified approximate
solutions in \S~\ref{sec:approx_temp}.  For an envelope in diffusive
equilibrium, \citet{fontaine.79} have presented an excellent
approximate solution for the composition profile.  In
\S~\ref{sec:approx_comp}, we extend their technique to the degenerate
regime.

\subsection{Analytic Temperature Profile}\label{sec:approx_temp}
Roughly speaking, the NS envelope can be divided into three zones, a
radiative outer zone, a sensitivity strip (where opacity changes from
primarily radiative to conductive), and a conductive isothermal
interior (also see \citealt{ventura.potekhin.01}).  In the radiative,
nondegenerate outer zone, the opacity is mostly determined by
free-free absorption. With an ideal gas equation of state, $yg = \rho
k_B T/\mu m_p$, where $\mu = A/(Z+1)$ is the mean molecular weight of
the plasma, the solution to the heat diffusion equation and
hydrostatic balance is
\begin{equation}\label{eq:thermal_nd_sol}
  T(y) = 1.58 \times 10^6 y^{4/17} \left(\mu Z
    g_{14} T_{e6}^4 \mu_e^{-2}\right)^{2/17} \Kelvin,
\end{equation}
where for brevity we have written $g_{14} = g/(10^{14} \CM\secs^{-2})$
and $T_{e6}=T_e/(10^6 \Kelvin)$.  For a radiative nondegenerate
envelope consisting of multiple layers of ions, the solution is
modified via the introduction of constants of integration at the
boundaries.  If the boundaries are sufficiently far away in pressure
from the sensitivity strip, the temperature profile of the underlying
layers will approximate that of a pure component.

In the sensitivity strip, which is defined by $\kappa_{\rm rad}(y_{\rm
  ss}) = \kappa_{\rm cond}(y_{\rm ss})$, the opacity transitions from
primarily radiative to primarily conductive. Here we follow the
formalism of \citet{ventura.potekhin.01}.  To find the sensitivity
strip, we take the approximate expression for
conductivity (\citealt{ventura.potekhin.01}),
\begin{equation}\label{eq:conductivity}
  K_c = 2.3 \times 10^{15} \frac {T_6} {\Lambda Z} \frac {\chi^3} { 1 +
    \chi^2} \frac {\rm ergs} {\CM\secs\Kelvin},
\end{equation}
where $\chi = p_F/m_ec \approx (\rho_6/\mu_e)^{1/3}$ is the relativity
parameter, $\mu_e = A/Z$, and $\Lambda$ is the Coulomb logarithm.  We
set $\Lambda = 1$ for simplicity's sake and $1 + \chi^2 \approx 1$
since the relativity parameter is small in the nonrelativistic regime.
Equating the two opacities, equations (\ref{eq:kramers_opacity}) and
(\ref{eq:conductivity}), we find that the sensitivity strip is defined
by the condition,
\begin{equation}
  y = 11.2 \frac{A^{1/3} T_6^{17/6} \mu_e^{2/3}} 
  {Z^{1/3} g_{14} \mu} \gsqcm.
\end{equation} 
Taking this condition and inserting the solution for the radiative
zone, we obtain expressions for the column depth of the sensitivity
strip, $y_{\rm ss}$,
\begin{equation}\label{eq:ss_parameters}
  y_{\rm ss} = 6.86 \times 10^4 \frac{Z T_{e6}^4 \mu_e}
  {g_{14}^2 \mu^2} \gsqcm,
\end{equation}
and temperature,
\begin{equation}
  T_{\rm ss} = 2.2 \times 10^7 \left(\frac{ Z^3 T_{e6}^{12}}
  {g_{14}^3 \mu^3}\right)^{2/17} \Kelvin,
\end{equation}
as a function of effective temperature, gravity, and composition.
Beyond the sensitivity strip ($y > y_{\rm ss}$), we solve the
constant flux equation using the conductive opacity,
\begin{equation}
  T \frac {dT} {dy} = 2.2 \times 10^{16} \frac{ Z T_{e6}^4 \mu_e}
  {\rho^2},
\end{equation}
when $1 + \chi^2 \approx 1$. Integration of this equation is
relatively straightforward given an equation of state.  There are two
cases that we need to consider.  For the nondegenerate regime, the
solution for $y > y_{\rm ss}$ is
\begin{equation}\label{eq:temp_cond_nd_sol}
  T = T_{\rm ss}\exp\left[-1.62 \times 10^4 \frac{Z \Lambda T_{e6}^4
      \mu_e}{g_{14}^2 \mu^2}\left(y^{-1} - y_{\rm
        ss}^{-1}\right)\right] \Kelvin,
\end{equation}
where we have joined it with the radiative solution at the sensitivity
strip. 

The electrons become more degenerate as the column increases.  We
define the degenerate boundary with our analytic degeneracy condition
(eq.  [\ref{eq:degenerate_cond}]), which gives the condition
\begin{equation}\label{eq:ydeg}
y = 22 \frac {T_6^{5/2} \Psi^{5/2} \mu_e^{5/3}} {g_{14} \mu^{5/2}} \gsqcm,
\end{equation}
where we have assumed that $y_{\rm deg} \gg y_{\rm ss}$.  Using our
solution for $T$ beyond the sensitivity strip (eq.
[\ref{eq:temp_cond_nd_sol}]), we find a system of
equations for $T_{\rm deg}$ and $y_{\rm deg}$, which are solved
numerically.

The solution beyond the degenerate boundary is
\begin{equation}\label{eq:temp_degenerate_sol}
  T^2 - T_{\rm deg}^2 = 1.37 \times 10^{16} T_{e6}^4 \frac Z
  {g_{14}^{6/5} \mu_e} \left(y_{\rm deg}^{-1/5} - y^{-1/5}\right) \Kelvin^2.
\end{equation}
Taking the limit of equation (\ref{eq:temp_degenerate_sol})
as $y \rightarrow \infty$ gives an analytic expression for the
core temperature, $T_c$, as a function of $T_e$, which reproduces
previous scalings (Gudmundsson, Pethick \& Epstein 1983; Potekhin,
Chabrier \& Yakovlev 1997).  \citet{ventura.potekhin.01} present a
more generalized solution in terms of the relativity parameter $\chi$.

In Figure \ref{fig:col_vs_conc_T} we present the approximate and
numerical solutions for the temperature and composition which we
discuss in the next section. For the thermal profile, the agreement is
excellent (to within 10\%).  It also agrees well with the analytic
solution given by \citet{ventura.potekhin.01}.

\subsection{Analytic Composition Profiles}\label{sec:approx_comp}
We find analytic solutions for the composition when one ion species is
dominant and the electrons are either nondegenerate or very
degenerate.  For the nondegenerate case, we first derive the result of
\citet{fontaine.79}. We start with the hydrostatic balance equations
for the ions, where ``1'' is the background and ``2'' is the trace,
\begin{eqnarray}
\frac {dP_1} {dr} = -n_1 \left( A_1 m_p g - Z_1 e E\right), \\
\frac {dP_2} {dr} = -n_2 \left( A_2 m_p g - Z_2 e E\right).
\end{eqnarray}
For the ions, $P_i = n_i k_B T$.  Dividing each pressure equation
above by their respective $P_i$ and subtracting ``1'' from ``2'', we
have:
\begin{equation}
k_B T\frac {d\ln\left(P_2/P_1\right)} {dr} = -\left(A_2 -
A_1\right)m_p g + \left(Z_2 - Z_1\right) e E,
\label{eq:conc_before_E}
\end{equation}
which in the trace limit ($n_2 \ll n_1$) gives
\begin{equation}
\ln\left(\frac {P_2} {P_1}\right) = \ln\left(\frac {n_2} {n_1}\right)
\approx \ln\left(\frac {n_2} {n_1 + n_2}\right) = \ln f_2,
\end{equation}
where $f_2$ is number fraction of the trace.

In the nondegenerate regime, the electric field is given by equation
(\ref{eq:rosseland_E}).  Using this result and using the trace
approximation so that $A \approx A_1$ and $Z \approx Z_1$, equation
(\ref{eq:conc_before_E}) becomes
\begin{equation}
  \frac {d\ln f_2} {dr} = -\frac {A_1 m_p g} {(Z_1 + 1) k_B T}
  \left[\frac {A_2} {A_1} (Z_1 + 1) - Z_2 - 1\right] 
\end{equation}
Using hydrostatic balance $dP/dr = -\rho g \approx -A_1 m_p g n_1$ and
$P \approx (Z_1 + 1) n_1 k T$, we have the elegant result,
\begin{equation}
  \frac {d \ln f_2} {d \ln P} = \left[\frac{A_2} {A_1} (Z_1 + 1) - Z_2 -
    1\right].
\label{eq:conc_nondeg}
\end{equation}
The concentration of a trace species is a power law of pressure with
an exponent, $\delta = A_2(Z_1 + 1)/A_1 - Z_2 - 1$.  It is widely
applicable because it gives the concentration in terms of a pressure
contrast,
\begin{equation}
f_2(P) = f_2(P_{\rm b}) \left(\frac P {P_{\rm b}}\right)^\delta,
\label{eq:sol_conc_nondeg}
\end{equation} 
where $f_2(P_{\rm b})$ is the concentration at an arbitrary boundary
$P_{\rm b}$. 

We now extend this result to the highly degenerate regime, where the
electron pressure dominates over ion pressure, and the electric field
is
\begin{equation}
eE = \frac {m_p g} {\mu_e} = \frac A Z m_p g.
\end{equation} 
Placing this into equation (\ref{eq:conc_before_E}) gives
\begin{equation}
\frac {d \ln f_2} {dP} = \frac {A_1 m_p} {\rho(P,T) k_B T(P)} \left[\frac {A_2}
{A_1} - \frac {Z_2} {Z_1}\right].
\end{equation}
We solve for $\rho(P,T)$ via the electron equation of state and $T(P)$
from the flux equation.  This solution is best when the electrons are
highly degenerate.  However, this is also the regime where the
temperature begins to be isothermal (\citealt{ventura.potekhin.01}),
allowing us to replace $T(P)$ by $T_{\rm deg}$, where $T_{\rm deg}$ is
the temperature at the degenerate boundary, giving
\begin{equation}
  \frac {d \ln f_2} {dP} = \frac {A_1 m_p} {\rho_{\rm deg}
    \left(P/P_{\rm deg}\right)^{1/\gamma} k_B T_{\rm deg}} \left[\frac
    {A_2} {A_1} - \frac {Z_2} {Z_1}\right],
\label{eq:degenerate_conc}
\end{equation}
where $\gamma$ is the polytropic index of the degenerate electron
equation of state, $P \varpropto \rho^{\gamma}$.  The condition of degeneracy (eq.
[\ref{eq:degenerate_cond}]) determines $P_{\rm deg}$ and $\rho_{\rm
  deg}$.

The electrons are nonrelativistic at the degenerate boundary so
$\gamma = 5/3$ and we integrate (eq. [\ref{eq:degenerate_conc}]) to
get
\begin{equation}
f_2 = f_{\rm 2,deg,0} \exp\left\{-\eta \frac {A_1 m_p P_{\rm deg}} {\rho_{\rm deg} k_B
  T_{\rm deg}}
  \left[\frac {Z_2} {Z_1} - \frac {A_2} {A_1}\right] \left[\left(\frac P
  {P_{\rm deg}}\right)^{\eta} - 1\right]\right\}
\label{eq:comp_deg_sol}
\end{equation}
where $f_{\rm 2,deg,0}$ is the concentration where the envelope
becomes degenerate and $\eta = 1 - \frac 1 \gamma = 0.4$.  Note that
$\eta$ is positive for all degenerate equations of state: the
concentration always decreases exponentially with increasing pressure.
Our solution reproduces the Boltzmann solution for a particle
experiencing a upward pointing force $F$ in the isothermal limit, $f_2
\varpropto \exp\left(F r/k T\right)$ since $r \varpropto P/\rho
\varpropto P^{\eta}$.  Thus, in the degenerate regime the
concentration falls faster than the power law relation in the
nondegenerate regime. In Figure \ref{fig:col_vs_conc_T}, we compare
the numerical solution with our analytic calculation. The agreement is
poor for higher densities due to the increasing importance of
relativistic equation of state, but gives reasonable agreement in the
burning layer $y_{\rm burn} \sim 10^{6} \tgsqcm$.

\section{Hydrogen Burning in Diffusive Equilibrium} \label{sec:hydrogen_burning}

The relevant nuclear processes that consume hydrogen depend on the
temperature and composition of the underlying matter.  We consider the
case of a NS which has a pure hydrogen layer on top of a layer of
proton capturing material (i.e. carbon, nitrogen, oxygen, etc) in
diffusive equilibrium.  We have not yet calculated the impact of an
intervening He layer, but we think it will be small as long as it does
not penetrate to a depth greater than the burning layer.

For simplicity, we first consider carbon. For hydrogen on carbon,
there are only two possible nuclear reactions: $\C({\rm
p},\gamma)\Nth$ and ${\rm p}({\rm p},e^+ +\nu_e){\rm D}$.  The D
produced in the p-p capture undergoes ${\rm D}({\rm
p},\gamma)\Hethree$ which in general is the endpoint of this process.
Since we always have a region where T $> 10^7 \Kelvin$, the reaction
rate for p-p capture is very slow compared to the proton capture rate
onto carbon.  The \nuclei{13}{N}\ decays to \nuclei{13}{C} which has
$A/Z = 13/6 \approx 2.167$, greater than the local $A/Z = 2$ for pure
carbon.  The $\Cth$ will thus sink through the outer layers of the NS,
and will not reside long enough in the burning layer to facilitate a
catalytic cycle.  Hence, the complete exhaustion of \Hyd\ by this
process requires an excess of \C\ compared to \Hyd.

\subsection{Diffusive Nuclear Burning}

We calculate the burning rate presuming that the burning time is slow
compared to the time for hydrogen to diffuse down through the carbon
to the burning layer.  We express this hierarchy of diffusion time
to nuclear burning time by studying the \Hyd\ equation of continuity,
\begin{equation}
\frac {\partial n_H} {\partial t} + \divr \J_H = -\frac {n_H}
{\tau_H},
\label{eq:continuity_full}
\end{equation} 
where $\tau_H = \langle\sigma v\rangle^{-1}n_C^{-1}$ is the local
lifetime of \Hyd\ to \C\ capture.  The condition of diffusive
equilibrium is $\partial n_H/\partial t = 0$ or more specifically that
$\partial n_H/\partial t$ changes only on the timescale associated
with depletion of the hydrogen column, which is long compared to both
the local nuclear burning time $\tau_H$ and the diffusion time and
therefore is dropped.  With the condition of steady state $\J_H$ and
$\JDNB$ are equivalent.  Writing the result in one dimension for $\J_H
= \JDNB = \D\partial n_H/\partial z - n_H v_{\rm dr}$, we have
\begin{equation}
\D \frac {\partial^2n_H} {\partial z^2} - v_{\rm dr} \frac {\partial
  n_H} {\partial z} = -\frac {n_H} {\tau_H},
\end{equation},
where $\D$ is the diffusion coefficient and $v_{\rm dr}$ is the drift
velocity.  We presume that $\D$ and $v_{\rm dr}$ change slowly
with $z$.  Our presumption that nuclear burning does not affect
diffusive equilibrium implies that locally the timescale associated
with nuclear burning is much longer than the timescale associated with
diffusion to the burning layer, or $\tau_H \gg \tau_{\rm ion} =
l^2/\D$, where $l$ is the ion scale height for \Hyd in the \C\
layer. Therefore, taking \D\ from \citet{brown.02}, we have
\begin{equation}\label{eq:taudr}
\tau_{\rm ion} = 800 \frac {T_6^{0.7} Z_i^{1.3}
        \rho_5^{0.6}} {A_i^{0.1} g_{14}^2 \left(A_i - Z_i\right)^2}
        \secs,
\end{equation}
where $A_i$ and $Z_i$ are the atomic number and charge of the
background proton capturing element.  

For proton capturing reactions, the proton lifetime
(\citealt{clayton}) is
\begin{equation}\label{eq:H_lifetime}
  \tau_{\rm H}^{-1} = 2.45 \times 10^{16} \left(\frac {S_0} {1\,{\rm
  keV}\,{\rm barns}}\right) \frac {\rho X_i} {A_i}
  T_6^{-2/3} \exp\left(-B T_6^{-1/3}\right) \yrs^{-1} 
\end{equation}
where $X_i$ and $A_i$ is the mass fraction and atomic number of the
proton capturing element and $S_0$ and $B$ are parameters determined
by the proton capturing element.  We derive an analytic expression for
the condition of diffusive equilibrium by expanding equation
(\ref{eq:H_lifetime}) about $T_6 = 40$,
\begin{equation}\label{eq:tauH}
\tau_H = 1.32 \times 10^5 \rho_5^{-1} \left(\frac {T_6} {40}\right)^{-14}
\secs.
\end{equation}
Comparing equations (\ref{eq:taudr}) and (\ref{eq:tauH}), our
condition for carbon is
\begin{equation}\label{eq:diff_burning_cond}
T_6 < 45.5 \rho_5^{-0.11},
\end{equation}
for $g_{14} = 2$.  

\begin{figure}[tbp]
  \epsscale{0.80} 
  
  \plotone{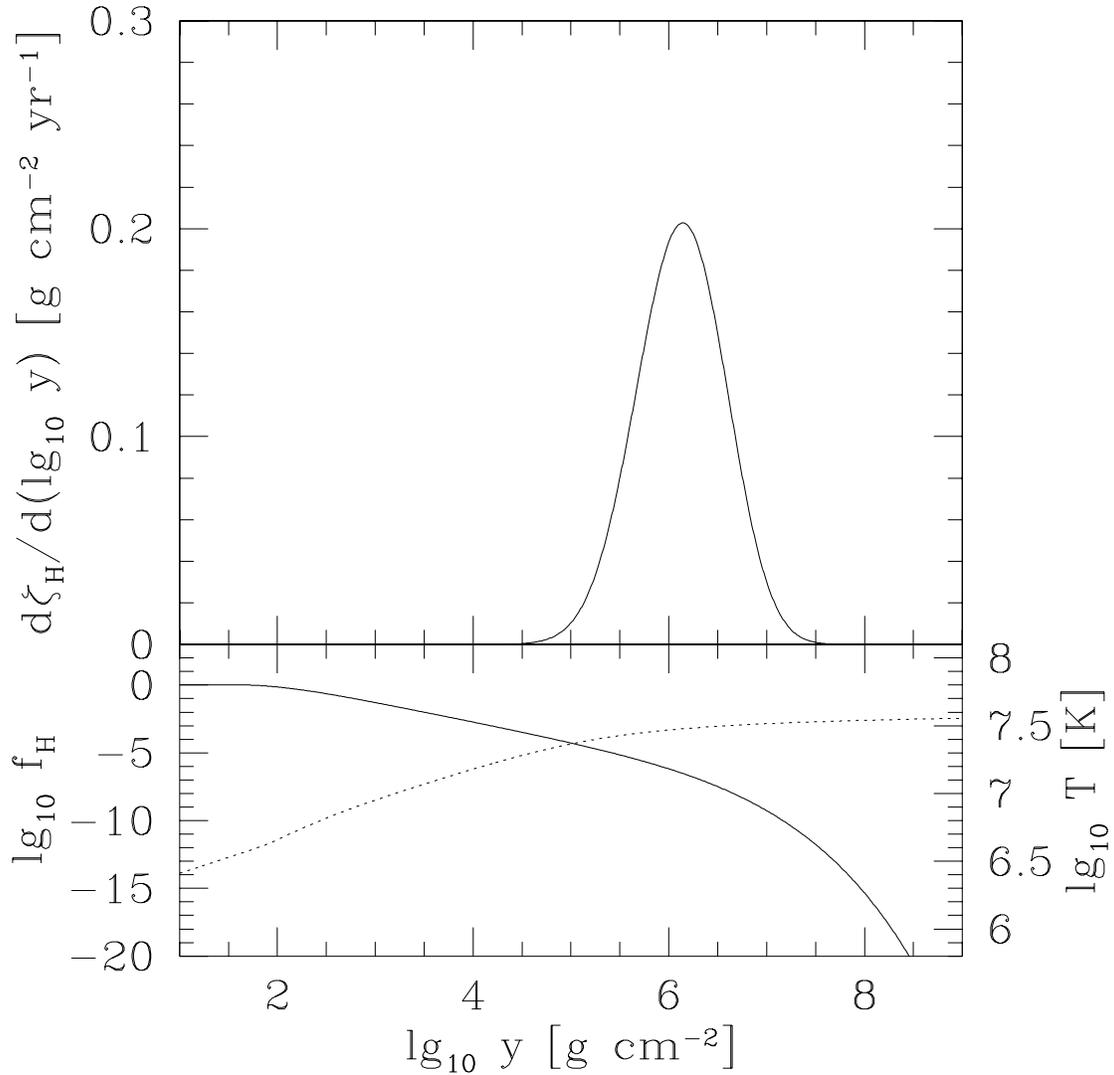}
  \caption{Differential hydrogen column burning rate taking into
    account p-p capture and p + \C\ capture.  The bottom graph shows
    the number fraction (solid line) and temperature (dotted line).
    This model has $y_H = 100 \tgsqcm$ and $T_e = 8 \times 10^5
    \Kelvin$.  The integrated burning rate for this model is
    $y_H/\tau_{\rm col} = 0.24 \tgsqcm\,{\rm yr}^{-1}$.  The burning
    peak occurs at a column of $y_{\rm burn} \approx 10^6 \tgsqcm$,
    where $T = 2.9 \times 10^7 \Kelvin$ and $\rho = 4.9 \times 10^4
    \tgccm$.  The local drift time ($\tau_{\rm ion} = 10^4\secs$) is
    much shorter than the local burning time ($\tau_H = 1.5 \times
    10^7 \secs$).}
    \label{fig:col_vs_burning}
\end{figure}

If the conditions for DNB are fulfilled for a sufficiently large
pressure range, the total burning rate of hydrogen, defined as
$\zeta_H = \JDNB({\rm H/\C\ boundary})$ is
\begin{equation}
\zeta_H = \frac {y_H} {\tau_{\rm col}} = 
\int \frac {n_H m_p} {\tau_H(n_H, n_C, T)} dz,
\label{eq:col_burning_rate}
\end{equation}
where $y_H$ is the integrated column of hydrogen, $\tau_{\rm col}$ is
the characteristic time for that column to be consumed and $n_H$ and
$n_C$ are the local number density of hydrogen and carbon
respectively.  This equation is incorporated into our flux and
structure equations and solved simultaneously, yielding $\tau_{\rm
  col}$ as a function of our stellar properties and $y_H$.  We plot
$d\zeta_H/d\lg_{10} y$ for an envelope with $y_H = 100 \tgsqcm$ in
Figure \ref{fig:col_vs_burning}. In the bottom graph we show the local
concentration and temperature.  Due to the inverse dependence of the
local hydrogen burning rate on temperature and concentration, the
burning layer is concentrated in a narrow pressure range at a depth
where the \Hyd\ abundance is small.  The diffusion timescale to this
depth is $\tau_{\rm ion} = 10^4\secs$, which is much shorter than the
nuclear timescale of $1.5 \times 10^7\secs$.  \emph{This observation is
  the essence of DNB, namely, that practically all the burning occurs
  in the exponentially suppressed diffusive hydrogen tail.}

\begin{figure}[tbp]
  \epsscale{0.80}
  \plotone{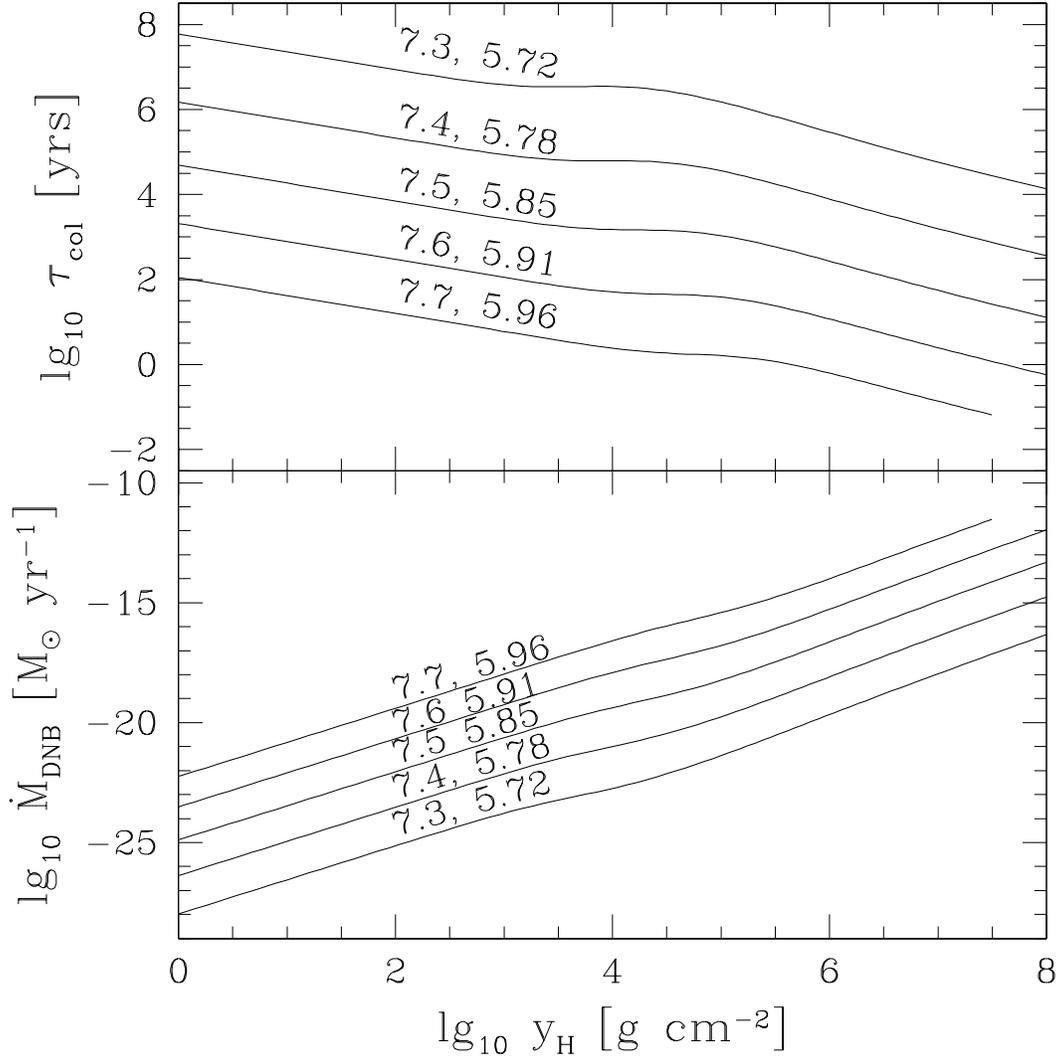}
    \caption{Lifetime of a hydrogen column, $\tau_{\rm col}$, and
      total mass burning rate, $\dot{M}_{\rm DNB}$, as a function of
      the size of the hydrogen column for different fixed core
      temperatures and $g_{14} = 2$.  For each model, we list the
      logarithmic core temperature and associated logarithmic
      effective temperature.  The bottom plot shows the total mass
      burning rate for a NS with a fiducial radius of $10 \,{\rm km}$
      for a hydrogen column of $y_H$.  For a central temperature of
      $lg_{10} T_c = 7.7$, our assumption of constant flux breaks down
      for columns greater than $3 \times 10^7\tgsqcm$.  At these large
      columns the heat release from nuclear burning becomes comparable
      to the flux. The power law dependence between the lifetime and
      $y_H$ is universal and the scaling does not change with
      different core temperatures.  The derivation of this and other
      power law scalings is described in \S~\ref{sec:approx_DNB}.  }
    \label{fig:col_burning}
\end{figure}

In Figure \ref{fig:col_burning}, we plot the characteristic time,
$\tau_{\rm col}$, and characteristic mass burning rate, $\dot{M}_{\rm
DNB} = 4\pi R^2 y_H/\tau_{\rm col}$, for $g_{14} = 2$ and fixed core
temperature, $T_c$.  For a central temperature of $5 \times 10^7
\Kelvin$, our calculations are valid only up to $3 \times 10^7
\tgsqcm$.  Above this column, the assumption of constant flux breaks
down since the energy released from DNB is comparable to the flux.

Two results are obvious from this figure.  First the relation between
lifetime (and hence mass burning rate) and column is related via a
simple power law between $\tau_{\rm col}$ and $y_H$.  We derive this
power law in \S~\ref{sec:approx_DNB}.  Secondly, the lifetime of the
entire envelope is set by the lifetime of the photosphere.

\begin{figure}[tbp]
  \epsscale{0.80}
  \plotone{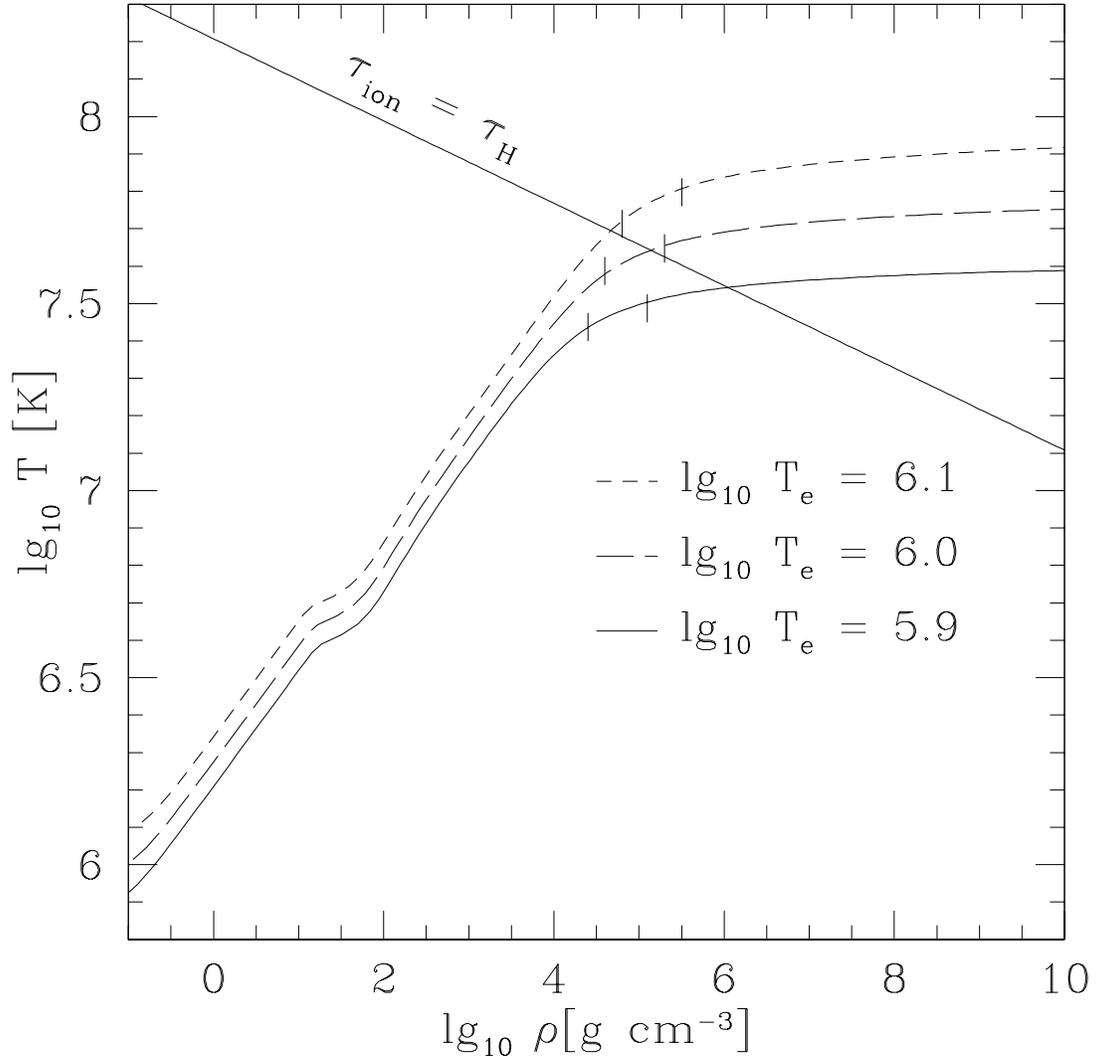} 
  
  \caption{Thermal structure of a NS with a total integrated \Hyd\
    column, $y_H \approx 100 \tgsqcm$ with different $T_{e,6} = (0.8,
    1, 1.26)$ respectively.  The burning layer for each of these
    models is represented by area between two vertical lines.  Also
    plotted is the line $\tau_{\rm ion} = \tau_H$, above which DNB
    does not occur in diffusive equilibrium.}
\label{fig:rho_vs_T_diff}
\end{figure}

The location of the burning layer must satisfy our derived diffusive
equilibrium condition, (eq. [\ref{eq:diff_burning_cond}]) for our
calculation to be valid.  In Figure \ref{fig:rho_vs_T_diff}, we plot
various models against our condition for DNB in diffusive equilibrium
(eq. [\ref{eq:diff_burning_cond}]).  The burning layers is the region
between the two vertical lines in each model. For models with $T_e >
10^6 \Kelvin$ (or $T_c > 6 \times 10^7 \Kelvin$), our present
calculation is not valid as DNB will occur in the slow diffusion
limit.  We will expand on this point and study the slow diffusion
limit in a future paper.

\subsection{Approximate Solution}\label{sec:approx_DNB}

The competition between the falling \Hyd\ abundance and rising
temperature with depth forces much of the hydrogen burning to occur in
a narrow zone.  As a result, we can solve the burning integral (eq.
[\ref{eq:col_burning_rate}]) over the burning layer by approximating
it with the method of steepest descents.

Since DNB occurs deep below the hydrogen layer, $X_i \approx 1$,
equation (\ref{eq:col_burning_rate}) then becomes
\begin{equation}\label{eq:col_burning_rate_int}
  \zeta_H = \frac {y_H} {\tau_{\rm col}} = 2.45 \times 10^{16}
  A_i^{-2} \left(\frac {S_0} {1\,{\rm keV}\,{\rm barns}}\right)\int
  {\rho} y f_H T_6^{-2/3} \exp\left(-B T_6^{-1/3}\right) d\ln(y)
  \yrs^{-1}
\end{equation}
where $f_H$ is the local number concentration of hydrogen, which
decreases with increasing $y$. Note that we have made a change of
variables of $dz = y \rho^{-1} d\ln(y)$.

This integral should be evaluated in both the nondegenerate, radiative
regime and the degenerate, conductive regime.  However, since the
majority of the burning occurs in the degenerate, conductive regime,
we will evaluate the integral there by the method of steepest
descents (\citealt{clayton}),
\begin{equation}
\int g(x)\exp(-h(x)) dx \approx g(x_0) \exp(-h(x_0)) \sqrt{\frac
  {2\pi} {h''(x_0)}},
\end{equation}
where $g(x)$ is a slowly varying function function of $x$, $h(x)$ is a
peaked function, and $x_0$ is the solution of $h'(x_0) = 0$, the
location of the peak of the burning rate.  In order to carry out this
approximation for equation (\ref{eq:col_burning_rate_int}), we make
the following identifications,
\begin{eqnarray}
  \exp(-h(x)) &\rightarrow& y \rho f_H \exp\left(-B T_6^{-1/3}\right),
  \nonumber\\
  g(x) &\rightarrow& T_6^{-2/3}.
\end{eqnarray}
For the degenerate, conductive regime, the peaked function is
\begin{equation}
-h\left[\ln(y)\right] = \frac 8 5 \ln(y) - B T_6^{-1/3}   
-\eta \frac {A_i m_p g y_{\rm deg}} {\rho_{\rm deg} k_B T_{\rm deg}}
  \left(\frac 1 {Z_i} - \frac 1 {A_i}\right) \left[\left(\frac y
  {y_{\rm deg}}\right)^{-\eta} - 1\right].
\end{equation}
The last term comes from the analytic solution for the composition
profile (eq. [\ref{eq:comp_deg_sol}]), where the background is the
proton capturing element and the trace is hydrogen. The temperature,
$T$, in the degenerate regime is given by equation
(\ref{eq:temp_degenerate_sol}).  Since the burning peak $y_0$ is
defined by a transcendental equation, $h'(\ln(y_0)) = 0$, we solve it
numerically and then calculate $h''(\ln(y_0))$.  For a total column of
hydrogen of $y_H = 100 \tgsqcm$, $T_e = 8 \times 10^5 \Kelvin$, and
$g_{14} = 2$, we find by the method of steepest descents, $\tau_{\rm
  col} = 468 \yrs$, which compares well with the numerical answer of
$\tau_{\rm col} = 428 \yrs$.  The burning layer is centered around
$y_0 = 7.2 \times 10^5 \tgsqcm$ with a local temperature of $T(y =
y_0) = 2.6 \times 10^7 \Kelvin$ and central core temperature $T_c =
3.3 \times 10^7 \Kelvin$.

Power law scalings of $\tau_{\rm col}$ with $y_H$, $T_e$ and $g$ can
also be determined from this integral.  The simplest power law scaling
involves $y_H$, which we determine as follows. The rate,
$y_H/\tau_{\rm col}$, is directly proportional to $f_H$ as in equation
(\ref{eq:col_burning_rate_int}).  From equation
(\ref{eq:sol_conc_nondeg}) and (\ref{eq:comp_deg_sol}), $f_H
\varpropto f_{\rm H,deg,0} = (y_{\rm deg}/y_H)^{\delta}$, where
$f_{\rm H,deg,0}$ is the hydrogen number fraction at the degenerate
boundary and $\delta = A_2(Z_1 + 1)/A_1 - Z_2 - 1 = -17/12$ for H on
\C.  Therefore, $\tau_{\rm col} \varpropto y_H^{1+\delta}$, which is
confirmed from comparisons to the numerical results.  This scaling is
accurate as long as $y_H < y_{\rm deg}$.

\begin{figure}[tbp]
  \epsscale{0.80} \plotone{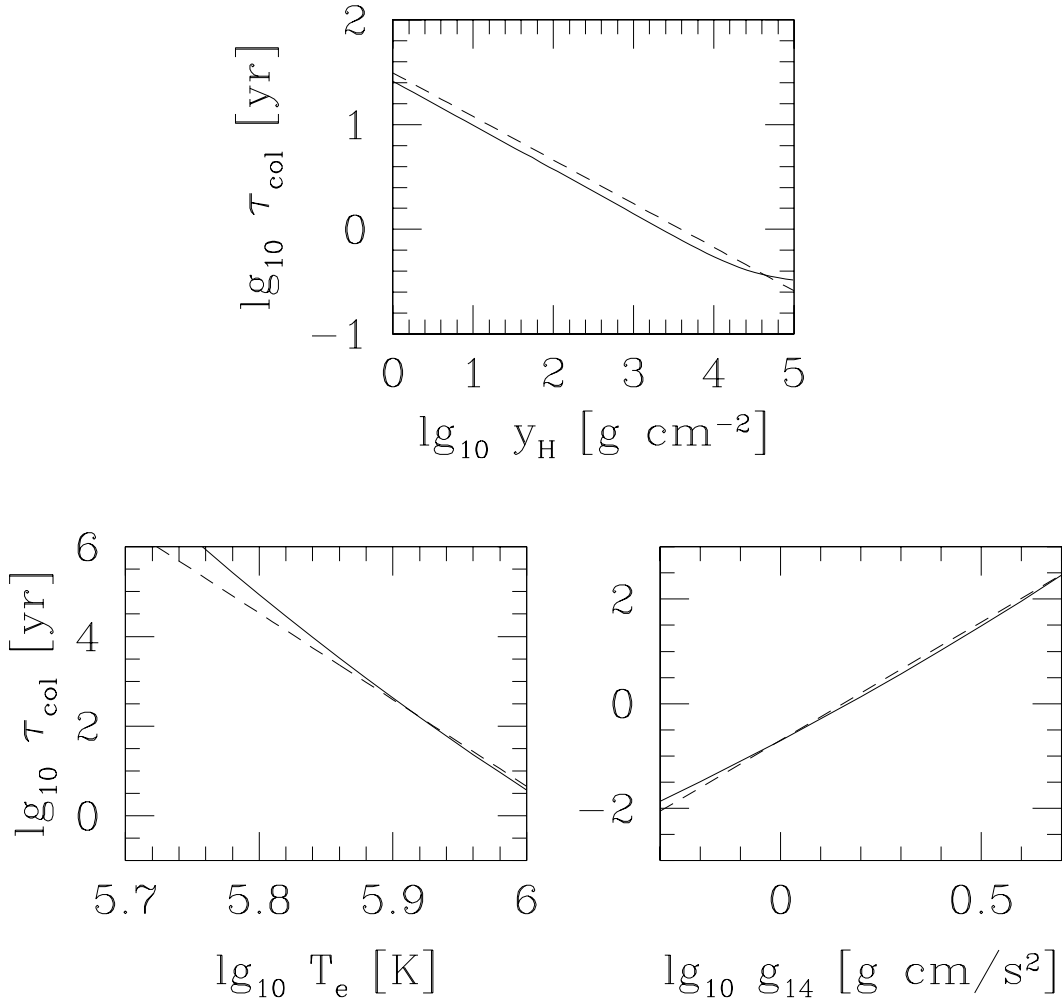}
  \caption{Comparison of scaling laws for the lifetime of the
  hydrogen column, $\tau_{\rm col}$.  The numerical solutions (solid
  lines) agree with a high degree of precision with the analytic
  scaling laws (dashed lines).  The comparison for effective
  temperature and local gravity is given for a fixed hydrogen column
  of $y_H = 100 \tgsqcm$.  For the plot of lifetime against total
  hydrogen column, the numerical calculation was done with a constant
  core temperature. The disagreement between analytic and numerical
  results for low effective temperatures is due to our chosen $T_{0,6}
  = 40$ for the power law expansion of the burning rate.  Expanding
  around a lower value of $T_{0,6}$ resolves this discrepancy.}
  \label{fig:g14_col_Te_vs_tau}
\end{figure}

The scaling of $\tau_{\rm col}$ against $T_{e6}$ and $g_{14}$ can be
derived from a power-law expansion of burning integral
(eq. [\ref{eq:col_burning_rate_int}]) in terms of $T$. The expansion of
the burning integral (eq. [\ref{eq:col_burning_rate_int}]) gives
\begin{equation}\label{eq:burning_integral_expansion}
  \frac {y_H} {\tau_{\rm col}} \varpropto f_H T_{0,6}^{-2/3}
  \exp\left(-B T_{0,6}^{-1/3}\right),
\end{equation}
where $T_{0,6} = T_0/10^6$ and $T_0$ is the temperature of the burning
layer.  Since we are in the degenerate conductive regime, $T_0 \approx
T_c$.  The location of the burning region, $y_0$ and $\rho_0$, scale
weakly with $T$ compared to the temperature of the burning region,
$T_0$.  Taking the analytic solution to the temperature profile (eq.
[\ref{eq:temp_degenerate_sol}]) and degeneracy point (eq.
[\ref{eq:ydeg}]), the power law dependence between $T_c$, $T_{e6}$ and
$g_{14}$ is $T_0 \varpropto T_c \varpropto T_{e6}^2 g_{14}^{-1/2}$.
Therefore, the scaling of $f_H$ in \C\ goes as,
\begin{equation}
  f_H \varpropto \exp\left\{-\eta \frac {m_p P_{\rm deg}} {\rho_{\rm deg} k_B T_0}
      \left[\left(\frac P {P_{\rm deg}}\right)^{-\eta} -
      1\right]\right\}.
\end{equation}
Expanding this equation at $T_0 = 10^8 K$ gives $f_H \varpropto
(T_0/10^8)^{0.475} g_{14}^{0.285} \varpropto T_{e6}^{0.95}
g_{14}^{0.0475}$.  Expanding the exponential in equation
(\ref{eq:burning_integral_expansion}) for carbon ($B = 136.96$) gives
$\exp\left( -B T_c^{-1/3}\right) \varpropto T_c^{9.8} \varpropto T_{e6}^{19.7}
g_{14}^{-4.9}$. Thus, the scaling for the rate is ${y_H}/{\tau_{\rm
    col}} \varpropto T_{e6}^{19.3} g_{14}^{-4.52}$.  Hence, the dominant
scaling comes from the exponential dependence on temperature of the
nuclear burning rate with small corrections from the other factors.
Putting all the scalings together with our analytic calculation, we
find
\begin{equation}\label{eq:Te_g14_vs_tau}
  \tau_{\rm col} \approx 6.5 \left(\frac{y_H}
  {100 \tgsqcm}\right)^{-5/12}T_{e6}^{-19.3} \left(\frac {g_{14}} 
  2\right)^{4.52} \yrs.   
\end{equation}
In Figure \ref{fig:g14_col_Te_vs_tau}, we compare the scaling found
numerically with this analytic result and find reasonable agreement.
The agreements for the scalings of $y_H$ and $g_{14}$ and the
numerical results are good.  However, the agreement for the scaling
of $T_e$ and numerical results diverge for low $T_e$.  This is due to
the changing power law with respect to expanding around different
values of $T_c$.  Expanding around a value of $T_c$ appropriate for
these lower values of $T_e$ resolves this discrepancy. 

These scalings can be calculated for other proton capturing elements.
The scaling of $\tau_{\rm col}$ and $y_H$ remains the same with the
exponent $1 - \delta_i$, where $\delta_i$ takes on different values
for the differing backgrounds $(Z_i,\ A_i)$.  We can use the same
strategy as earlier, expanding our results in $T_0$ and inserting the
scalings for $T_{e6}$ and $g_{14}$ afterwards.  Hence, $f_H$ has the
same scaling as before, $f_H \varpropto (T_0/10^8)^{0.475}
g_{14}^{0.285}$.  However, the dominant scaling from the exponential
dependence of temperature on the burning rate changes. Hence we choose
to ignore the scalings associate with $f_H$.  In terms of the
exponential factor $B_i$, this dependence becomes $\exp\left( -B_i
T_{0,6}^{-1/3}\right) \varpropto
\left({T_0}/{10^8\Kelvin}\right)^{B_i/14} \varpropto T_{e6}^{B_i/7}
g_{14}^{-B_i/28}$. Since $f_H$ does not scale strongly with $T_0$
compared to the nuclear reaction rate (especially with heavier
elements where $B_i$ gets progressively larger), we ignore the scaling
of $f_H$ with $T_0$ and $g_{14}$.  To put this another way, the value
of the number density, $f_H$, at the burning layer does not change as
drastically as the value of the nuclear
timescale at the burning layer with $T_0$.  Putting these
scalings all together, we have
\begin{equation}\label{eq:power_law_tau}
\tau_{\rm col_i} = \tau_{\rm col_i,0} \left(\frac {y_H}
  {100 \tgsqcm}\right)^{1+\delta_i} T_{e6}^{-B_i/7} g_{14}^{B_i/28},
\end{equation} 
Table \ref{table:tau_values} lists the values of the scalings, the
prefactor, calculated both numerically and analytically, and the
largest effective temperature where DNB in diffusive equilibrium is
valid.  Above $T_{e,6\,\mathrm{DNB}}$, equation
(\ref{eq:power_law_tau}) is not valid since DNB no longer occurs in
diffusive equilibrium.

\subsection{Application to Cen X-4 in Quiescence}\label{sec:cenx-4}

The large number of transiently accreting NS also provide an
astrophysical site for our work. The work of Rutledge et al. (1999)
(see Bildsten and Rutledge 2001 for an overview) first showed that
much of the quiescent emission from these objects was thermal emission
from the surface, allowing for many $T_e$ measurements
(\citealt{rutledge._cen_x4,rutledge._aql_x1, rutledge.00,
  rutledge.99}). Of all these LMXBs, only Cen X-4 is safely in the DNB
regime.  Its effective temperature is low enough ($T_{e} = 8.8 \times
10^5$ \citealt{rutledge._cen_x4}) so that DNB in diffusive equilibrium
is a good approximation for a hydrogen column $y_H < 10^8 \tgsqcm$.
The column depth of H on \cen\ after an outburst is expected to be
$y_H < 10^8 \tgsqcm$; set by ignition conditions of type I X-ray
bursts (\citealt{brown.02}).

Accretion outbursts were observed in 1969 and 1979 from \cen, but the
latter outburst was particularly weak.  Presuming no outburst has
occurred in the intervening years, this gives an age of 23 years. For
a hydrogen column of $y_H = 10^8\tgsqcm$ after the 1979 outburst, we
find that the present parameters of \cen\ would be $y_H \approx 2
\times 10^6 \tgsqcm$ and $T_c \approx 4 \times 10^7$ K if there has
been no accretion in quiescence. A much larger H column ($y_H \approx
2 \times 10^{7} \tgsqcm$) is needed to burn matter via DNB at the
supplied rate if the NS is accreting in quiescence at a rate
$\dot{M}_q = 10^{-13} \Msun\,{\rm yr}^{-1}$.

\begin{figure}[tbp]
  \plotone{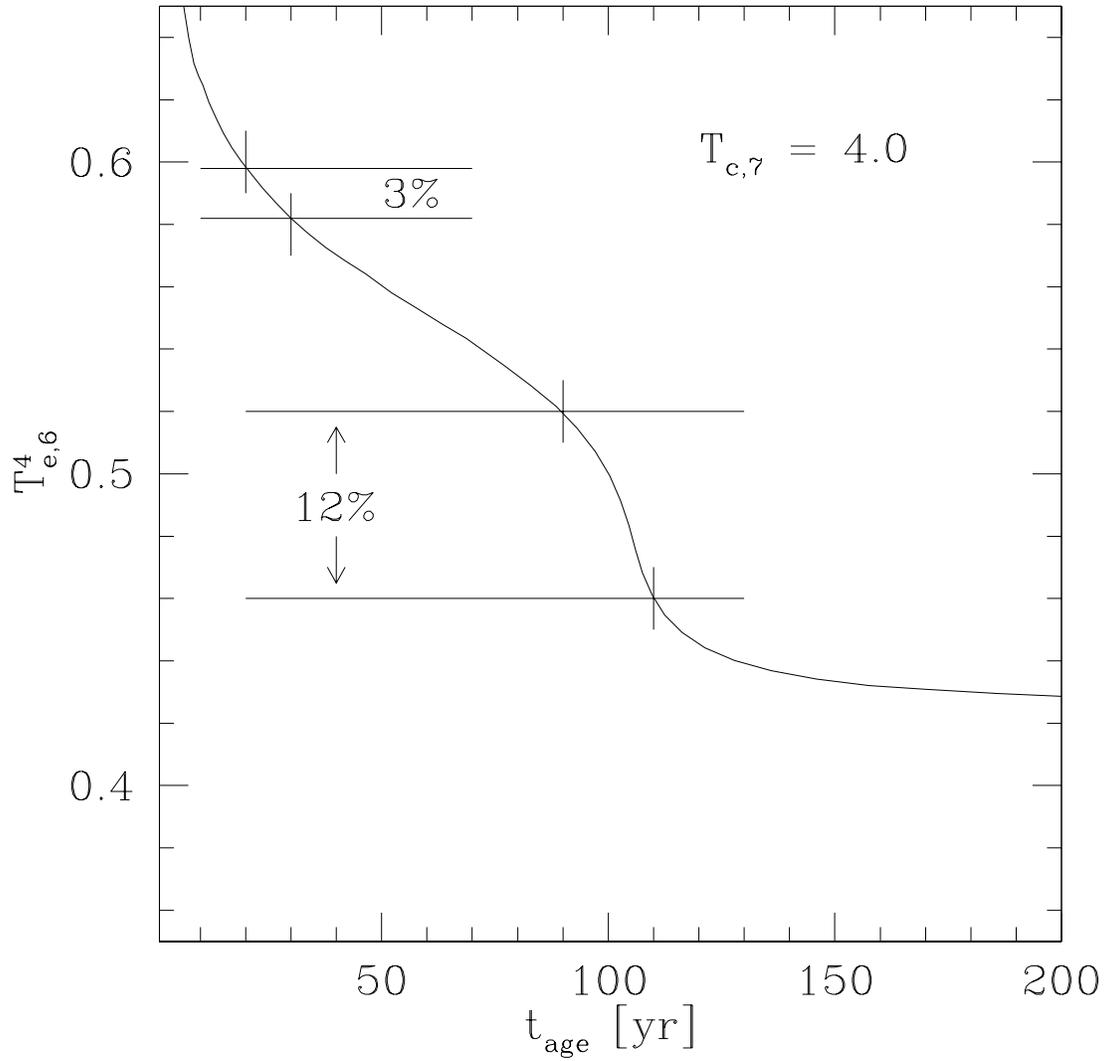} \caption{Flux as a function of the age
    of the \Hyd/\C\ envelope for a core temperature of $T_c = 4 \times
    10^7$ \Kelvin.  The $\sim 12\%$ drop in flux over a twenty year
    period at 100 \yrs after an outburst is indicative of DNB.  Also
    shown is the $3\%$ variation over a ten year period at the present
    time.}  \label{fig:cen_tau_vs_Teff}
\end{figure}

Presuming little accretion in quiescence and fixed $T_c$, we evolve
$y_H$ and the resulting flux as a function of the time since the last
outburst.  The age of the envelope or time since the last outburst,
$t_{\rm age}$ is the age derived from the scaling of $\tau_{\rm col}$
with $y_H$ in the analytic solution,
\begin{equation}
t_{\rm age} = \int \left(\frac {dy_H} {dt}\right)^{-1} dy_H = \frac
{12} {5} \tau_{\rm col}.
\end{equation}
In Figure \ref{fig:cen_tau_vs_Teff}, we plot the resultant flux from a
total integrated column of \Hyd, $y_H$, against the lifetime of that
column for fixed core temperature, $T_c$.  Evolution via DNB predicts
that the outgoing flux will drop by $\sim 12\%$ over a twenty year
period roughly 100 years after the outburst.  This drop is due to the
consumption of \Hyd\ that makes up the sensitivity strip of the
\emph{\Hyd\ layer}.  At present (twenty years after the outburst) the
flux will vary by only 3\% over a ten year timescale.  The associated
$\dot{M}_{\rm DNB} = 2 \times 10^{-15} \Msun\,{\rm yr}^{-1}$ for this
time after the outburst.  Observations of \cen\ have placed
constraints on the variation of $T_e$ over a five year timescale of $<
10\%$ (\citealt{rutledge._cen_x4}).  Observational confirmation of DNB
on \cen\ is therefore unlikely.
    
\section{Conclusions and Discussions}\label{sec:discussion}

We have shown that diffusive nuclear burning (DNB) is an effective
mechanism that burns surface hydrogen on a NS on astrophysically
relevant timescales.  Our numerical and analytic solutions for the
rates of DNB are in the limit where the H is in diffusive equilibrium
with the underlying proton capturing elements.  Our work is similar to
that originally carried out by Michaud and Fontaine
(\citealt{michaud.84a,michaud.84b}) for diffusion-induced hydrogen
burning in white dwarfs and also by
\citet{iben.85a,iben.85b}. However, there are several crucial
differences: the NS temperatures are much higher and the local gravity
much stronger.  Hence, we were able to make simplifying assumptions
and avoid the active diffusion calculation with nuclear burning
performed by \citet{iben.85a}. We have also solved for the
concentration profile in the degenerate regime self-consistently,
whereas the effects of degeneracy were neglected in Michaud and
Fontaine's paper.

The largest remaining uncertainty in applying our work is the unknown
composition of the material underlying the hydrogen.  Our assumptions
of low magnetic fields and surface temperatures ($T_e<10^6 \ {\rm K}$
for DNB to be nuclear-rate limited) limit the DNB applications to
quiescent NSs in transient low-mass X-ray binaries and millisecond
radio pulsars. Little is known about the surface composition of
millisecond radio pulsars, though in the absence of our work one would
definitely expect H to be dominant on the surface since these objects
underwent mass accretion prior to becoming a pulsar. If there is
enough carbon in the underlying composition from previous H/He burning
during type I X-ray bursts (Schatz et al. 1999), then there is clearly
adequate time to burn off the H. A clear detection of H at the
photosphere of a millisecond radio pulsar would thus constrain the
underlying abundances.

The impact of an intervening helium layer depends on its thickness.  A
thin layer of helium, which does not penetrate into the burning layer
and remains in the nondegenerate regime, enhances the burning rate.
This is because the electric field in a nondegenerate helium plasma
($E = 4 m_p g/3e$) is smaller than the electric field in a
nondegenerate carbon plasma ($E = 12 m_p g/7e$).  Hence the diffusive
tail of hydrogen can penetrate more easily into the helium layer onto
the underlying layer of proton capturing elements and the number
fraction of hydrogen is enhanced at the burning layer.  For a thick
helium layer which goes beyond the burning layer, DNB would be
effectively shut off.  However, a thick helium layer would not likely
sit on top of a thick proton capturing layer, if the proton capturing
elements have the same A/Z as helium and both layers are degenerate.
This is because the electric field is no longer a differentiating
factor allowing the helium to mix downward with the proton capturing
material.  This would effectively dilute the abundance of proton
capturing elements, thereby reducing the rate, but not shutting off
DNB. The timescale for this mixing and its exact impact on the overall
burning rate still needs more careful attention.

The recent observations of a pair of X-ray spectral lines on
1E1207.4-5209 (\citealt{sanwal.1e1207,mereghetti.1e1207}) show that
hydrogen is not present on the surface of this young NS
(\citealt{sanwal.1e1207,hailey.mori.2002}).  These spectral lines
appear to be mid-atomic elements like oxygen or neon
(\citealt{hailey.mori.2002}) or helium if $B \sim 10^{14} \Gauss$
(\citealt{sanwal.1e1207}).  Since the age of the associated SN remnant
is 7 kyrs, the mechanism of hydrogen removal must be extremely fast.
However, our calculation is not directly applicable to this
system. The spindown of 1E1207.4-5209 implies a dipole B-field of $B
\approx 3 \times 10^{12} \Gauss$ (\citealt{pavlov.1e1207}).  This
object also has a high temperature of $T \approx 1.4-1.9\,{\rm MK}$ at
which our assumption of diffusive equilibrium breaks down. However,
the physics remains the same and we expect this mechanism to be active
in this system.

Since we have explored one special limit of DNB, it is not surprising
that we have not found a physical system in which DNB is directly
observable.  In a future paper we will consider the problem of
diffusion limited DNB, which is applicable to systems where the
assumption of diffusive equilibrium breaks down ($T_e > 10^6\Kelvin$
for \Hyd/\C).  We will also consider the problem of DNB in highly
magnetized sources such as young radio pulsars ($B \sim
10^{12-13}\Gauss$) and magnetars ($B \sim 10^{14-15}\Gauss$).

\acknowledgements

   We thank D. Uzdensky for showing us a nice mathematical trick, and
R. Sunyaev for reminding us of the importance of the magnetic field on
the thermal structure in radio pulsars.  We would also like to thank
S. W. Davis, C. J. Deloye, A. Socrates and D. Townsley for discussions
and the referee for important clarifications. P.C. would like to thank
the Department of Physics and Department of Astronomy at Columbia
University, where the early and latter parts of this work was done,
for their hospitality.  This research was supported by NASA via grant
NAG 5-8658 and by the NSF under Grants PHY99-07949 and
AST01-96422. L. B. is a Cottrell Scholar of the Research Corporation.

\clearpage

\bibliographystyle{apj}
\bibliography{master}

\begin{thebibliography}{31}
\expandafter\ifx\csname natexlab\endcsname\relax\def\natexlab#1{#1}\fi

\bibitem[{{Alcock}(1980)}]{alcock.80a}
Alcock, C. 1980 \apj, 242, 710

\bibitem[{{Alcock} \& Illarionov(1980) Alcock \& Illarionov}]{alcock.80b}
Alcock, C. \& Illarionov, A. 1980 \apj, 235, 534

\bibitem[{Althaus \& Benvenuto(2000)}]{althaus.2000}
Althaus, L.~G. \& Benvenuto, O.~G. 2000, \mnras, 317, 952

\bibitem[{{Baiko} {et~al.}(1998)}]{baiko.ea.98}
Baiko, D.A., Kaminker, A.D., Potekhin, A.Y. \& Yakovlev, D.G. 1998, \apj, 81, 25

\bibitem[{Bildsten} {et~al.}(1992)]{bildsten.92}
Bildsten, L., Salpeter, E.~E. \& Wasserman, I. 1992, \apj, 384, 143

\bibitem[{{Bildsten} {et~al.}(1998)Bildsten \& Cumming}]{bildsten.cumming.98}
Bildsten, L. \& Cumming, A. 1998 \apj, 506, 842

\bibitem[{Bildsten \& Rutledge(2001)}]{bildsten.01}
Bildsten, L. \& Rutledge, R.~E. 2002, in The Neutron Star--Black Hole
  Connection, ed. C.~Kouveliotou, J.~Ventura, \& E.~{van den Heuvel}, Vol. 567,
  NATO ASI sec.~C (Dordrecht: Kluwer), 245, astro-ph/0005364


\bibitem[Brown et~al.(2002)]{brown.02}
Brown, E.F., Bildsten, L. \& Chang, P. 2002, \apj, 574, 920

\bibitem[{{Chapman} {et~al.}(1952) Chapman \& Cowling}]{chapman.52}
Chapman, S. \& Cowling, T.~G. 1952, \textit{Mathematical Theory of Non-Uniform Gases} (Cambridge: Cambridge University Press)

\bibitem[{Chiu \& Salpeter(1964)}]{chiu.64} 
Chiu, H.~Y., Salpeter, E.~E., 1964, \prl, 12, 413
 
\bibitem[{{Clayton}(1983)}]{clayton}
{Clayton}, D. 1983 \textit{Introduction to Stellar Structure and Nucleosynthesis} (Chicago: University of Chicago Press)

\bibitem[{Fontaine \& Michaud(1979)}]{fontaine.79}
Fontaine, G. \& Michaud, G. 1979, \apj, 231, 826

\bibitem[{{Gudmundsson} {et~al.}(1983){Gudmundsson}, {Pethick}, \&
  {Epstein}}]{gudmundsson83}
{Gudmundsson}, E.~H., {Pethick}, C.~J., \& {Epstein}, R.~I. 1983, \apj, 272,
  286

\bibitem[Hailey \& Mori(2002)]{hailey.mori.2002}
Hailey, C.~J. \& Mori, K. 2002, \apj, 578, L133

\bibitem[Hernquist \& Applegate(1984)]{hernquist.applegate:NS_analytic_calculation}
Hernquist, L. \& Applegate, J.~H. 1984, \apj, 287, 244
 
\bibitem[{Iben \& MacDonald(1985a)}]{iben.85a}
Iben, I. \& MacDonald, J. 1985, \apj, 296, 540

\bibitem[{Iben \& MacDonald(1985b)}]{iben.85b}
Iben, I. \& MacDonald, J. 1985, \apj, 301, 164

\bibitem[{{Itoh} {et~al.}(1983){Itoh}, {Mitake}, {Iyetomi}, \&
  {Ichimaru}}]{itoh83}
{Itoh}, N., {Mitake}, S., {Iyetomi}, H., \& {Ichimaru}, S. 1983, \apj, 273, 774

\bibitem[{{Itoh} {et~al.}(1985){Itoh}, {Nakagawa}, \&
  {Kohyama}}]{itoh85:_relat}
{Itoh}, N., {Nakagawa}, M., \& {Kohyama}, Y. 1985, \apj, 294, 17

\bibitem[{Itoh \& Kohyama(1993)}]{itoh93}
Itoh, N. \& Kohyama, Y. 1993, \apj, 404, 268

\bibitem[{Macdonald} {et~al.}(1998)]{macdonald._efield}
Macdonald, J., Hernanz, M. \& Jose, J. 1998, \mnras, 296, 523

\bibitem[Mereghetti et~al.(2002)]{mereghetti.1e1207} Mereghetti, S.,
De Luca, A., Caraveo, P.~A., Becker, W., Mignani, R., Bignami, G.~F.,
to appear in \apj, astro-ph/0207296

\bibitem[{Michaud {et~al.}(1984)}]{michaud.84a}
Michaud, G., Fontaine, G. \& Charland, Y. 1984, \apj, 280, 247

\bibitem[{Michaud \& Fontaine(1984)}]{michaud.84b}
Michaud, G. \& Fontaine, G. 1984, \apj, 283, 787

\bibitem[{Paczynski(1983)}]{paczynski.83}
{Paczynski}, B. 1983, \apj, 267, 315

\bibitem[Pavlov, Zavlin \& Sanwal(2002)]{pavlov:review_chandra}
Pavlov, G.~G., Zavlin, V.~E. \& Sanwal, D. in Proceedings of the
270. Heraeus Seminar on Neutron Stars, Pulsars and Supernova Remnants,
Bad Honnef (Germany), Jan.  21-25, 2002, eds. W. Becker, H. Lesch and
J. Truemper, astro-ph/0206024

\bibitem[Pavlov et~al.(2002)]{pavlov.1e1207}
 Pavlov, G.~G., Zavlin, V.~E., Sanwal, D., Trumper, J. 2002, \apj, 569, L95

\bibitem[{{Potekhin} {et~al.}(1997){Potekhin}, {Chabrier}, \&
  {Yakovlev}}]{potekhin97}
{Potekhin}, A.~Y., {Chabrier}, G., \& {Yakovlev}, D.~G. 1997, \aap, 323, 415

\bibitem[{{Potekhin} {et~al.}(1999){Potekhin}, {Baiko}, {Haensel}, \&
  {Yakovlev}}]{potekhin.ea.99}
{Potekhin}, A.~Y., {Baiko}, D.~A., {Haensel}, P., \& {Yakovlev}, D.~G. 1999,
  \aap, 346, 345

\bibitem[Potekhin \& Yakovlev(2001)]{potekhin.yakovlev:magnetized.envelopes}
{Potekhin}, A.~Y. \& {Yakovlev},D.~G. 2001, \aap 374, 213

\bibitem[Rajagopal \& Romani(1996)]{rajagopal.96}
Rajagopal, M. \& Romani, R.~W. 1996, \apj, 461, 327

\bibitem[{Rosen(1968)}]{rosen.68}
  Rosen, L.~C., 1968, \apss, 1, 372

\bibitem[{Rutledge {et~al.}(1999)}]{rutledge.99}
Rutledge, R.~E., Bildsten, L., Brown, E.~F., Pavlov, G.~G. \& Zavlin, V.~E. 1999, \apj, 514, 945

\bibitem[{Rutledge {et~al.}(2000)}]{rutledge.00}
Rutledge, R.~E., Bildsten, L., Brown, E.~F., Pavlov, G.~G. \& Zavlin, V.~E. 2000, \apj, 529, 985

\bibitem[{Rutledge {et~al.}(2001a)}]{rutledge._cen_x4}
Rutledge, R.~E., Bildsten, L., Brown, E.~F., Pavlov, G.~G. \& Zavlin, V.~E. 2001, \apj, 551, 921

\bibitem[{Rutledge {et~al.}(2001b)}]{rutledge._aql_x1}
Rutledge, R.~E., Bildsten, L., Brown, E.~F., Pavlov, G.~G. \& Zavlin, V.~E. 2001, \apj, 559, 1054
\bibitem[Sanwal et~al.(2002)]{sanwal.1e1207}
Sanwal, D., Pavlov, G.~G., Zavlin, V.~E., Teter, M.~A. 2002, \apj, 571, L61

\bibitem[Schatz et~al.(1999)]{schatz.99}
{Schatz}, H., {Bildsten}, L., {Cumming}, A., \& {Wiescher}, M. 1999, \apj, 524,
  1014

\bibitem[{Urpin \& Yakovlev(1980)}]{urpin80}
Urpin, V.~A. \& Yakovlev, D.~G. 1980, \sovast, 24, 126

\bibitem[{Ventura \& Potekhin(2002)}]{ventura.potekhin.01}
Ventura, J. \& Potekhin, A.~Y. 2002, in The Neutron Star--Black Hole
  Connection, ed. C.~Kouveliotou, J.~Ventura, \& E.~{van den Heuvel}, Vol. 567,
  NATO ASI sec.~C (Dordrecht: Kluwer), 393, astro-ph/0104003

\bibitem[Woosley \& Weaver(1995)]{woosley.95}
Woosley, S.~E. \& Weaver, T.~A. 1995, \apjs, 101, 181

\bibitem[{Yakovlev {et~al.}(2001)}]{yakovlev.01}
Yakovlev, D.G., Kaminker, A.D., \& Gnedin, O.Y. 2001, \aap, 379, 5

\bibitem[{Zavlin {et~al.}(1996)}]{zavlin.96}
Zavlin, V.~E., Pavlov, G.~G., Shibanov, Y.~A. 1996, \aap, 315, 141
 
\end{thebibliography}

\clearpage

\begin{deluxetable}{c c c r r r r}
  \tablecolumns{6}
  \tablewidth{0pt}
  \tablecaption{Power law exponents and prefactors and range of
    validity for equation (\ref{eq:power_law_tau}).
    \label{table:tau_values}} 

  \tablehead{\colhead{Reaction} &
    \multicolumn{2}{c}{$\lg_{10}\left(\tau_{\rm col_i,0}\right)$
      [yrs]} & \colhead{$B_i$} &
    \colhead{$1 + \delta_i$} & \colhead{$T_{e,6\,\mathrm{DNB}}$
      [\Kelvin]} \\
    \colhead{} &
    \colhead{numerical} & \colhead{analytic} & \colhead{} &
    \colhead{} & \colhead{}}
        
        \startdata
        \C(p,$\gamma$)\Nth & 0.57 & 0.58 & 136.93 & -5/12 & 1\\
        \N(p,$\gamma$)\nuclei{15}{O} & 2.02 & 2.15 & 152.31 & -6/14 &
        1.2 \\
        \Oxygen(p,$\gamma$)\nuclei{17}{F} & 3.06 & 3.23 & 166.96 &
        -7/16 & 1.3 \\
\enddata
\end{deluxetable}


\end{document}